\documentclass[aps,floatfix]{revtex4}
\usepackage[pdftex]{color,graphicx}

 \newcommand{\be}{\begin{equation}}
 \newcommand{\ee}{\end{equation}}
 \newcommand{\bea}{\begin{eqnarray}}
 \newcommand{\eea}{\end{eqnarray}}

\renewcommand\P{{\cal P}}

\begin{document}

\begin{flushright}
arXiv:1005.4053v2 [astro-ph.CO]\\
YITP-10-26
\end{flushright}

\title{Large-scale Perturbations from the Waterfall Field in Hybrid Inflation}

\author{Jos\'e Fonseca$^{1,2}$, Misao Sasaki$^{2}$ and David Wands$^{1,2}$}

\affiliation{{}$^{1}$ Institute of Cosmology and Gravitation, Dennis Sciama Building, Burnaby Road,
University of Portsmouth, Portsmouth PO1~3FX, United Kingdom\\
and\\
{}$^{2}$ Yukawa Institute for Theoretical Physics, Kyoto University, Kyoto 606-8502, Japan}

\begin{abstract}
We estimate large-scale curvature perturbations from isocurvature fluctuations in the waterfall field during hybrid inflation, in addition to the usual inflaton field perturbations. The tachyonic instability at the end of inflation leads to an explosive growth of super-Hubble scale perturbations, but they retain the steep blue spectrum characteristic of vacuum fluctuations in a massive field during inflation. The power spectrum thus peaks around the Hubble-horizon scale at the end of inflation. We extend the usual $\delta N$ formalism to include the essential role of these small fluctuations when estimating the large-scale curvature perturbation. The resulting curvature perturbation due to fluctuations in the waterfall field is second-order and the spectrum is expected to be of order $10^{-54}$ on cosmological scales.
\end{abstract}
\maketitle

%\tableofcontents

\section{Introduction}

Cosmological inflation driven by a single scalar field has many
attractively simple properties. In particular the quantum
fluctuations of a slow-rolling scalar field have an almost scale
invariant power spectrum and these give rise to adiabatic curvature
perturbations on super-Hubble scales which have an effectively
Gaussian distribution for canonical scalar fields. This is
consistent with current observations of primordial curvature
perturbations, but there is much interest in possible
deviations from this simple paradigm which would require either
more than one relevant field during inflation or a non-canonical
field, or both. Multiple fields can support non-adiabatic
perturbations on super-Hubble scales which can affect the
resulting primordial power spectrum and lead to non-Gaussianity.

Hybrid inflation, driven by the energy density of a false
vacuum state which is destabilised when a slow-rolling field
reaches a critical value, was originally proposed
by Linde \cite{Linde:1991km} and subsequently analysed by Linde
\cite{Linde:1993cn} and many others
\cite{Mollerach:1993sy,Copeland:1994vg}.
It has proved to be particularly successful for realising
inflation in supersymmetric models of particle physics \cite{Copeland:1994vg,Dvali:1994ms,Lyth:1998xn,Mazumdar:2010sa}.
In hybrid models there is
another field, usually called the waterfall field, which is
trapped in a false vacuum state until the instability is
triggered. Nonetheless the vacuum fluctuations of this field are
usually neglected since the field is massive when scales
corresponding to large-scale structure of our observable universe
leave the horizon. Quantum fluctuations of a massive scalar field,
with mass much larger than the Hubble scale, remain overdamped
even on super-Hubble scales, corresponding to decaying,
oscillating mode functions and thus a steep blue power spectrum.

However it was recently suggested
\cite{Mulryne:2009ci,Levasseur:2010rk} that non-adiabatic
large-scale perturbations in the waterfall field could play an
important role, in particular leading to non-Gaussian curvature
perturbations \cite{Mulryne:2009ci}, since these long wavelength
modes experience the most rapid growth due to the tachyonic
instability in the waterfall field at the end of inflation. It is
known that a slow transition during hybrid inflation, allowing
inflation to continue for some period after the tachyonic
instability, could lead to large curvature perturbations on scales
which leave the horizon around the time of the instability
\cite{Randall:1995dj,GarciaBellido:1996qt}. But it is difficult to
model the large-scale primordial curvature perturbations through the
phase transition where perturbations about the classical
background necessarily become large.

We reconsider this issue in this paper, using a
$\delta N$-formalism~\cite{Sasaki:1995aw,Sasaki:1998ug,Wands:2000dp,Lyth:2004gb}
to evaluate the primordial curvature perturbation on
large scales. We stress the essential role of small, Hubble-scale
field perturbations at the end of inflation in determining the
local integrated expansion $N=\int H\, dt$ in parts of the universe with
different values of the waterfall field averaged on large,
super-Hubble scales. We show that the variance of the waterfall
field becomes dominated by Hubble-scale perturbations when the
tachyonic instability begins, rapidly leading to the end of
inflation. The duration of inflation is shown to be independent of
the large-scale field at first-order, simply due to the symmetry
of the potential. The curvature perturbation due to long-wavelength modes of the waterfall field are shown to be suppressed due to the steep blue spectrum of the waterfall field fluctuations, similar to the case of false vacuum inflation
supported by thermal corrections~\cite{Gong:2008ni}.

\section{Waterfall field perturbations}
\label{SecPert}

We will consider the original hybrid inflation model~\cite{Linde:1991km,Linde:1993cn,Copeland:1994vg} which is
described by a slowly rolling inflaton field, $\phi$, and the waterfall
field, $\chi$, with a potential energy
\be \label{pot}
V(\phi,\chi)=\bigg(M^2-\frac{\sqrt{\lambda}}{2}\chi^2\bigg)^2+\frac{1}{2}m^2\phi^2+\frac{1}{2}\gamma\phi^2\chi^2
\ee The first term in the potential is a Mexican hat potential for
the waterfall field with the false vacuum at $\chi=0$ and true
vacuum at $\chi^2=2M^2/\sqrt{\lambda}$ when $\phi=0$. The
effective mass of the waterfall field in the false vacuum state is
 \be
 m_\chi^2(\phi)
 % = \gamma \phi^2 - 2\sqrt\lambda M^2
  = \gamma \left( \phi^2 - \phi_c^2 \right) \,.
 \ee
where we define $\phi_c^2\equiv 2\lambda^{1/2}M^2/\gamma$.
Thus the false vacuum is stabilised for $\phi^2>\phi_c^2$, while
for $\phi^2<\phi_c^2$ there is a tachyonic instability.
%Without loss of generality we take $\phi>0$ before the instability in the following.

Note that the simple potential (\ref{pot}) with a real scalar
field, $\chi$, has two discrete minima at $\chi=\pm
2M/\sqrt{\lambda}$. Thus regions which settle into different true
vacuum states are separated by domain walls at late times. However
vacuum states with higher-dimensional vacuum manifolds may have
cosmic strings, monopoles or no topologically stable defects. We
will neglect the formation of cosmic defects while noting their
presence could have important cosmological consequences in
particular hybrid models~\cite{Copeland:1994vg,GarciaBellido:1996qt}.

\subsection{Background solution}

In a spatially-flat Friedmann-Lemaitre-Robertson-Walker (FLRW)
cosmology the evolution equations for the background fields are
\bea
\ddot{\phi}+3H\dot{\phi}&=&-(m^2+\gamma\chi^2) \phi \label{eqmphi}\\
\ddot{\chi}+3H\dot{\chi}&=&(2\sqrt{\lambda}M^2-\gamma\phi^2-\lambda\chi^2)\chi \label{eqmchi}
\eea
where the Fridmann equation gives the Hubble rate
\be \label{Fried}
H^2=\frac{8\pi G}{3}\bigg( V(\phi,\chi)+\frac{1}{2}\dot{\phi}^2+\frac{1}{2}\dot{\chi}^2\bigg)
\ee

Initially we assume $\phi>\phi_c$ so that the $\chi$ field is held
in the false vacuum and we have the background solution $\chi = 0$.
For simplicity we will work in the vacuum-dominated regime
\cite{Copeland:1994vg} where we can neglect the energy density of the
$\phi$ field upon the Hubble expansion, $H\approx$constant,
 \be
 \label{deS}
a = H_c^{-1} \exp \left[H_c\left(t-t_c\right)\right]
 \quad {\rm where}\ H_c^2=\frac{8\pi G M^4}{3} \,.
 \ee
Then the late-time solution for Eq.~(\ref{eqmphi}) yields
 \be \label{solphi}
\phi = \phi_c \exp{\left[-rH_c\left(t-t_c\right)\right]}
 \ee
with
 \be
 \label{defr}
r=\frac{3}{2}-\sqrt{\frac{9}{4}-\frac{m^2}{H_c^2}} \,.
 \ee
The $\phi$ field slow-rolls for $m\ll H$ until the critical point, $t=t_c$, when the tachyonic instability is triggered. In our numerical examples we choose $r=0.1$. One can verify that this is consistent with vacuum-dominated regime (\ref{deS}) so long as $\gamma$ is not extremely small \cite{Copeland:1994vg}.
%This happens due to the fact for small $\phi$ the waterfall field gets a negative mass square. It happens when $\phi=\phi_*$ with %$\phi_*\equiv\sqrt{(\beta/\gamma)}H$. We choose $t=0$ to correspond to the beginning of the transition. We also scale the scale factor such that %$a(t=0)H=1$, then $\phi=\phi_*$ corresponds to $\eta=-1$, where $\eta$ is conformal time defined in Eq. (\ref{confdef}).

\subsection{Linear perturbations}

Quantum fluctuations in the slow-rolling $\phi$ field (coupled to
scalar metric perturbations) lead to an almost scale-invariant
spectrum of adiabatic curvature perturbations on super-Hubble scales
during inflation \cite{GarciaBellido:1996ke}. These correspond to local perturbations in the
evolution along the classical $\chi=0$ background solution. They
give rise to effectively Gaussian primordial curvature perturbations
usually considered in hybrid inflation models
\cite{Copeland:1994vg}.
By contrast quantum fluctuations in the waterfall field correspond
to isocurvature field perturbations during inflation decoupled from both
$\phi$-field and metric perturbations at linear order
\cite{Gordon:2000hv}.

Linear $\chi$-field perturbations, with comoving wavenumber $k$, obey the evolution equation
 \be
\ddot{\delta\chi_k}+3H\dot{\delta\chi_k}+\bigg(\frac{k^2}{a^2}-\beta H_c^2+\gamma\phi^2\bigg) \delta\chi_k = 0 \,.
 \label{lineardeltachi}
 \ee
where the bare tachyonic mass of the waterfall field relative to the inflationary Hubble scale is given by
 \be
\beta=2\sqrt{\lambda}\frac{M^2}{H_c^2} \,.
 \ee
We assume $\beta\gg1$ so that the time-scale associated with the tachyonic instabilty is much less that a Hubble time and inflation ends soon after the instability begins. If $\beta$ is of order unity or less then there is the possibility of an extended period of slow-roll inflation and associated large metric perturbations on scales associated with the transition \cite{Randall:1995dj,GarciaBellido:1996qt}. In our numerical solutions we choose $\beta=100$.

If we substitute the solution for $\phi$ (Eq.(\ref{solphi})), and rewrite the evolution equation (\ref{lineardeltachi}) in term of the canonically quantised variable, $u=a\delta\chi$, then we obtain a wave equation with time-dependent mass
 \be \label{upert}
u''_k + \left[k^2+\mu^2(\eta) \right]u_k=0
\quad {\rm where}\ \mu^2(\eta) \equiv \frac{\beta (|\eta|^{2r}-1)-2}{\eta^2} \,.
 \ee
where primes denote derivatives with respect to conformal time, $\eta=\int dt/a$. Note that in de Sitter (\ref{deS}) the conformal time is given by
 \be \label{confdef}
\eta = -\frac{1}{aH_c} = - \exp \left[-H_c\left(t-t_c\right)\right] \,.
 \ee
We have chosen the normalisation for the scale factor in Eq.~(\ref{deS}) such that $aH=1$ and $\eta=-1$ when $t=t_c$, i.e., Fourier modes with $k=1$ leave the Hubble horizon when $\phi=\phi_c$ and the tachyonic instability is triggered.

%We solve the evolution equation for the field perturbations of the waterfall field $\chi$. We fixed $\beta=100$ and $r=0.1$ and solved the previous equation for different $k$. Initially the modes are taken to be in a Bunch-Davis vacuum. In Fig. \ref{logu2} we plot the square of the amplitude of 5 different modes. One can see that these decrease their amplitude until the transition is attained. Once the mode crosses the potential barrier it starts growing in a k-independent manner as described in (\ref{lateutend}). This shown in the plot by the dashed black line.

\begin{figure}
\centering
\includegraphics[width=0.9\textwidth]{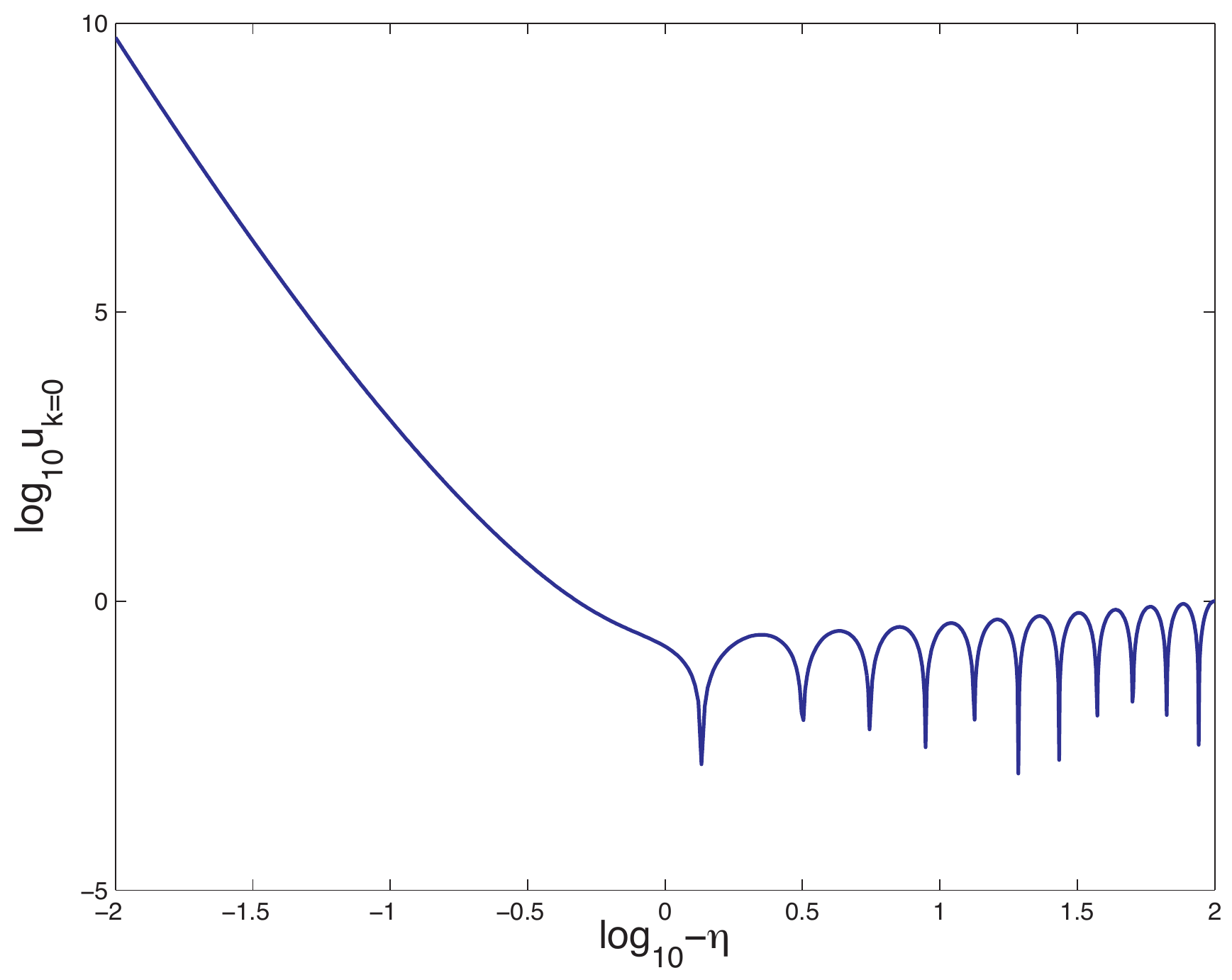}
\caption{The amplitude of the infinite
wavelength solution, $u_0$ given in Eq.~(\ref{u0}), plotted as a
function of conformal time, $\eta$, for $r=0.1$ and $\beta=100$.}
\label{uk0}
\end{figure}

For $k=0$ we find an analytic solution~\cite{GarciaBellido:1996qt}
 \be
 \label{u0}
u_0 = (-\eta)^{1/2} \left[ C J_{-\nu}\left(\frac{\sqrt\beta (-\eta)^r}{r}\right) + D J_{\nu}\left(\frac{\sqrt\beta (-\eta)^r}{r}\right) \right] \,,
\ee
where $r$ is defined in Eq.~(\ref{defr}) and
\be
 \nu = \frac{\sqrt{\frac{9}{4}+\beta}}{r} \,.
\ee
In Figure~\ref{uk0} we show a typical solution for $u_0(\eta)$.

\begin{figure}
\centering
\includegraphics[width=0.9\textwidth]{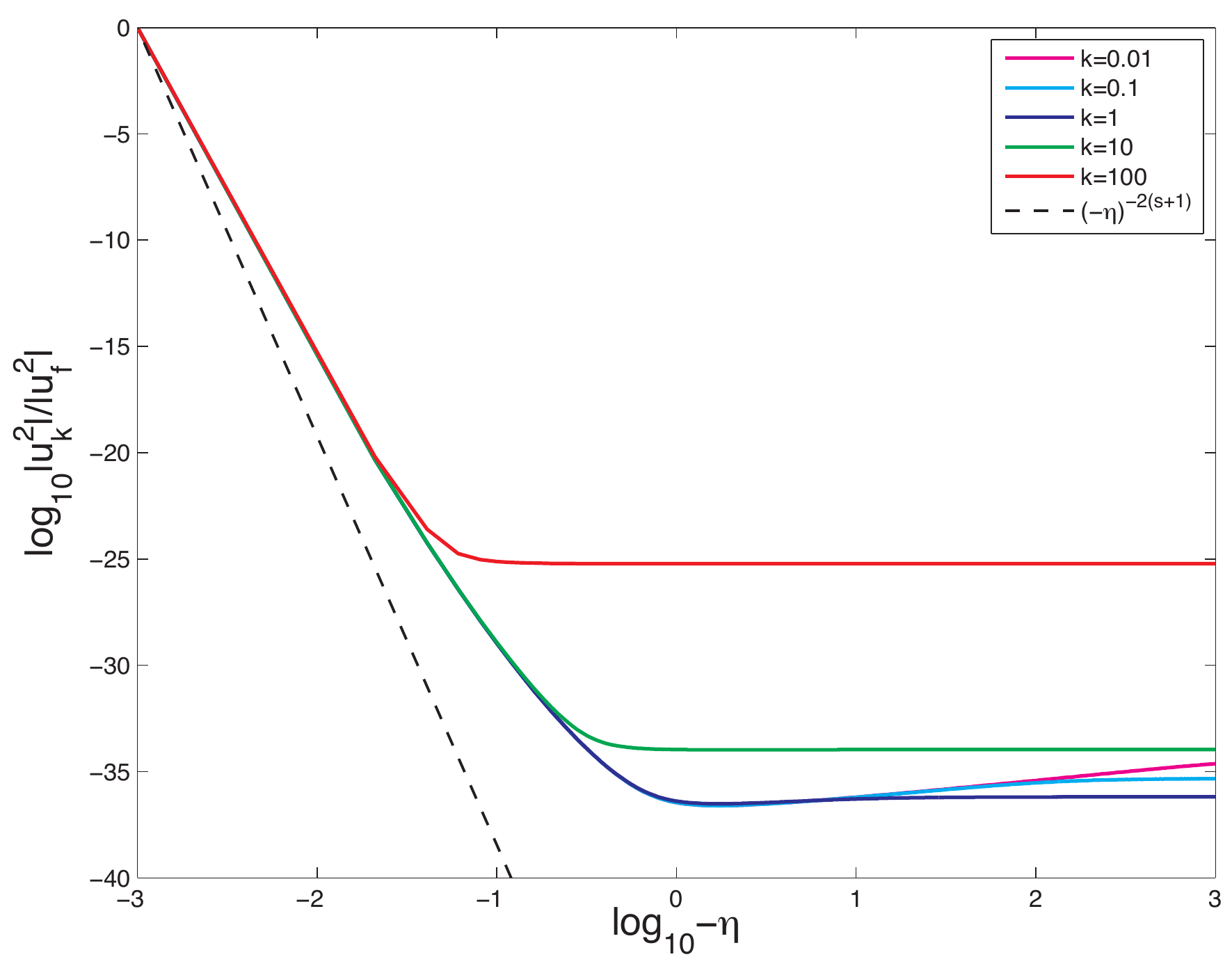}
\caption{The amplitude of the mode functions, $|u_k^2|$, normalized with respect to their final value, as a function of conformal time, $\eta$, for $r=0.1$ and $\beta=100$. $k=1$ corresponds to modes leaving the Hubble-horizon when $\phi=\phi_c$. The black dashed line indicates the expected late time behaviour, Eq.~(\ref{lateutend}), as $\phi\to0$.}
\label{logu2}
\end{figure}

For all modes with a finite $k$ the time-dependent mass-squared in
Eq.~(\ref{upert}) is positive,
$\mu^2(\eta)\to\beta|\eta|^{-2(1-r)}$, but negligible at early times
where $k\eta\to-\infty$. We thus normalise the mode functions at
early time to the quantum vacuum for a free field, $u_k\propto
e^{-ik\eta}/\sqrt{2k}$.
The time-dependent mass-squared reaches a maximum
 \be
 \mu^2_{\rm max} = \frac{(\beta+2)r}{1-r} \left( \frac{\beta(1-r)}{\beta+2} \right)^{1/r} \sim \beta r
 \quad {\rm when}\ \eta_{\rm max} = - \left( \frac{\beta+2}{\beta(1-r)} \right)^{1/2r} \sim -1 \,,
 \ee
and then tends to minus infinity at late times $\mu^2(\eta)\to
-(\beta+2)/\eta^2$ as $\eta\to0$. At late times we have the
asymptotic solution
 \be
  \label{lateutend} u_k \propto (-\eta)^{-s-1}
  \ee
corresponding to the tachyonic growing mode solution $\chi
=\chi_c\exp{[sH(t-t_c)]}$ with
 \be
  \label{s} s = \sqrt{\frac{9}{4}+\beta}-\frac{3}{2} \,.
 \ee
The amplitudes of the mode functions, $|u_k^2(\eta)|$ for different values of $k$ are shown in Figure~\ref{logu2}.

The power spectrum is defined as
 \be
  {\cal P}_u=\frac{k^3}{2\pi^2}\left|u_k^2\right| \,.
 \ee
Modes with $k\ll 1$ leave the Hubble-horizon well before the
instability, when $|\eta|\gg1$. At these early times the
time-dependent mass term in Eq.~(\ref{upert}) is positive with
$\mu^2(\eta)\simeq \beta|\eta|^{-2(1-r)}\gg k^2$, so that the mode
function is suppressed on super-Hubble scales and we find
$|u^2|\propto |\eta|^{2(1-r)} \propto a^{-2(1-r)}$. Normalising the
initial state to the quantum vacuum for a free field then leads to
a steep blue spectrum on super-Hubble scales with ${\cal
P}_\chi\propto k^3$, which is even steeper than the quantum vacuum
for a free field on the smallest scales, ${\cal P}_\chi\propto
k^2$. There are no classical perturbations on super-Hubble scales
before the transition.

\begin{figure}
\centering
\includegraphics[width=0.9\textwidth]{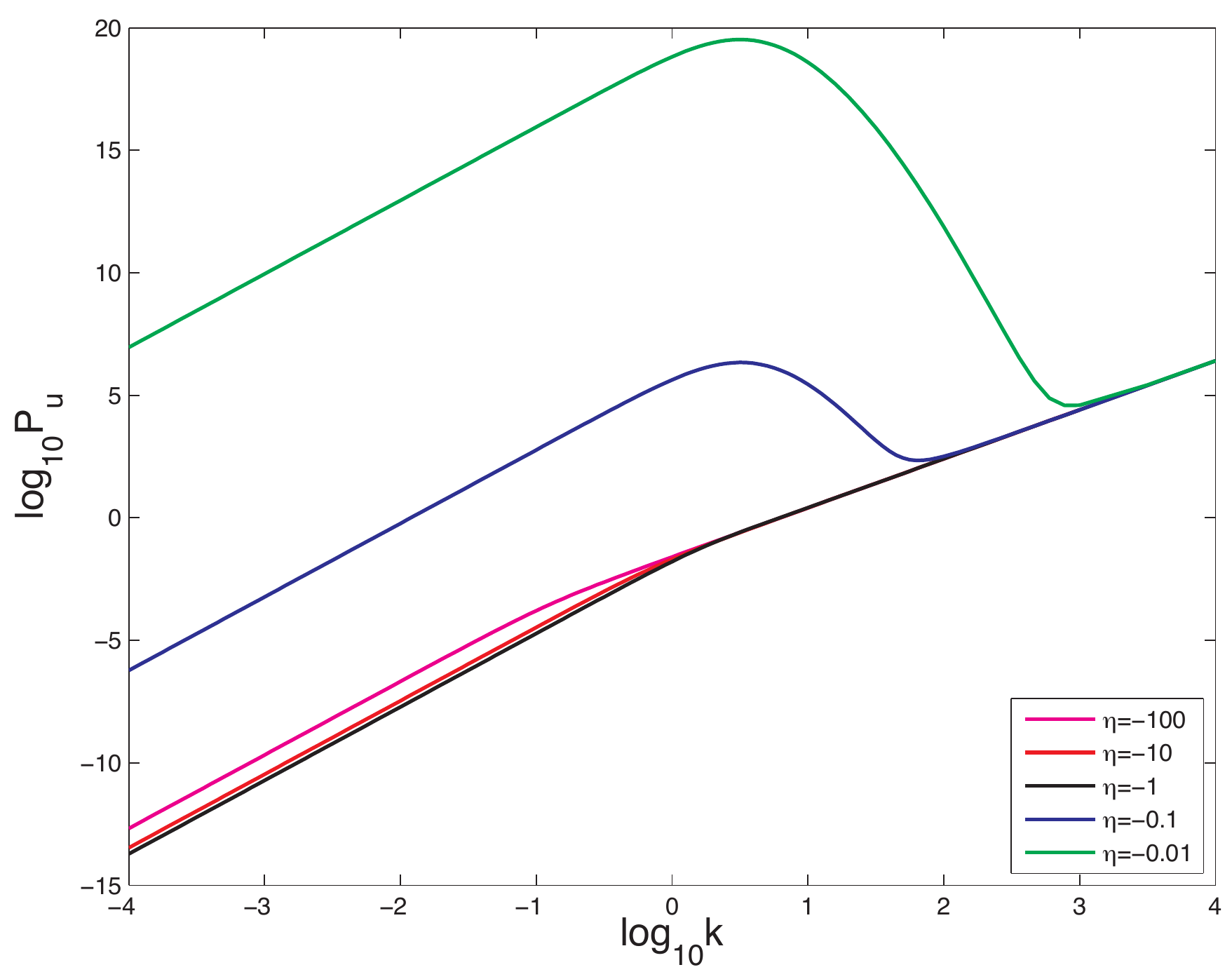}
\caption{The power spectrum of $u_k$ as a function of
the comoving wavenumber, $k$, relative to the Hubble-horizon at the
transition. We show the power spectrum at four different times to
show its time evolution. On large scales, $k\ll 1$ the power is suppressed before the transition, $\eta<\eta_*$, but then grows rapidly due to the tachyonic instability, for $\eta>\eta_*$, where $\eta_*\approx -1$.} \label{pu2}
\end{figure}

Modes with $k\gg1$ remain within the Hubble horizon until after
the tachyonic instability is triggered. Even after $\phi=\phi_c$,
gradient terms stabilise short-wavelength modes with
$k^2+\mu^2(\eta)>0$ and longer-wavelength modes begin to grow first. As
a result we find that the power spectrum for the waterfall field
begins to peak on scales of order the Hubble horizon at the
transition, $k\sim 1$. This is clearly shown in Figure~\ref{pu2}, where we
fix $\beta=100$ and $r=0.1$.

%\begin{figure}
%\centering
%\includegraphics[width=0.9\textwidth]{logPuvslogkdifetasmooth.pdf}
%\caption{ \scriptsize The vertical axis we have the logarithm of the power spectrum of $u_k$, $\mathcal{P}_{u_k}$. On the horizontal axes we plot the logarithm of the k-modes. We plot the power spectrum for different times to see its time evolution.}
%\label{pu2}
%\end{figure}

\section{Application of $\delta N$ formalism}

In order to calculate the primordial curvature perturbation, $\zeta$, due to fluctuations in the waterfall field we can calculate the perturbed expansion, $\zeta \equiv \delta N\equiv \delta (\ln a)$ \cite{Starobinsky:1986fxa,Sasaki:1995aw,Sasaki:1998ug,Wands:2000dp,Lyth:2004gb}, from an initial spatially-flat hypersurface to a final uniform-density hypersurface \cite{Malik:2008im}. For adiabatic perturbations in the $\phi$-field during the slow-roll phase this yields the standard first-order result
 \be
 \zeta_\phi = - \frac{H\delta\phi}{\dot\phi} \,.
 \ee
where $H\delta\phi/\dot\phi$ can be evaluated on spatially-flat hypersurfaces any time after Hubble exit for each Fourier mode.
Perturbations in the waterfall field are non-adiabatic field perturbations and hence we must follow their effect on the expansion through the tachyonic transition from false to true vacuum at the end of inflation. At first order we can simply add these independent contributions to the primordial curvature perturbation from both fields
 \be
 \zeta = \zeta_\phi + \zeta_\chi \,.
 \ee

The usual ``separate universe'' approach \cite{Wands:2000dp} is to
calculate the classical expansion for different initial values of
the field assuming a locally homogeneous and isotropic (i.e.,
FLRW) cosmology, assuming the long-wavelength behaviour is
independent of much shorter wavelength modes. However in the
hybrid inflation model the classical solution assuming homogeneous
fields is liable to give an incorrect estimate of the duration of
the transition. Indeed the classical background trajectory assumed
in section~\ref{SecPert} was $\chi=0$ for which, classically,
inflation never ends and $N=\ln(a)\to\infty$ as $t\to\infty$ and
$\phi\to0$ in Eqs.~(\ref{deS}) and (\ref{solphi}).

\begin{figure}
\centering
\includegraphics[width=0.9\textwidth]{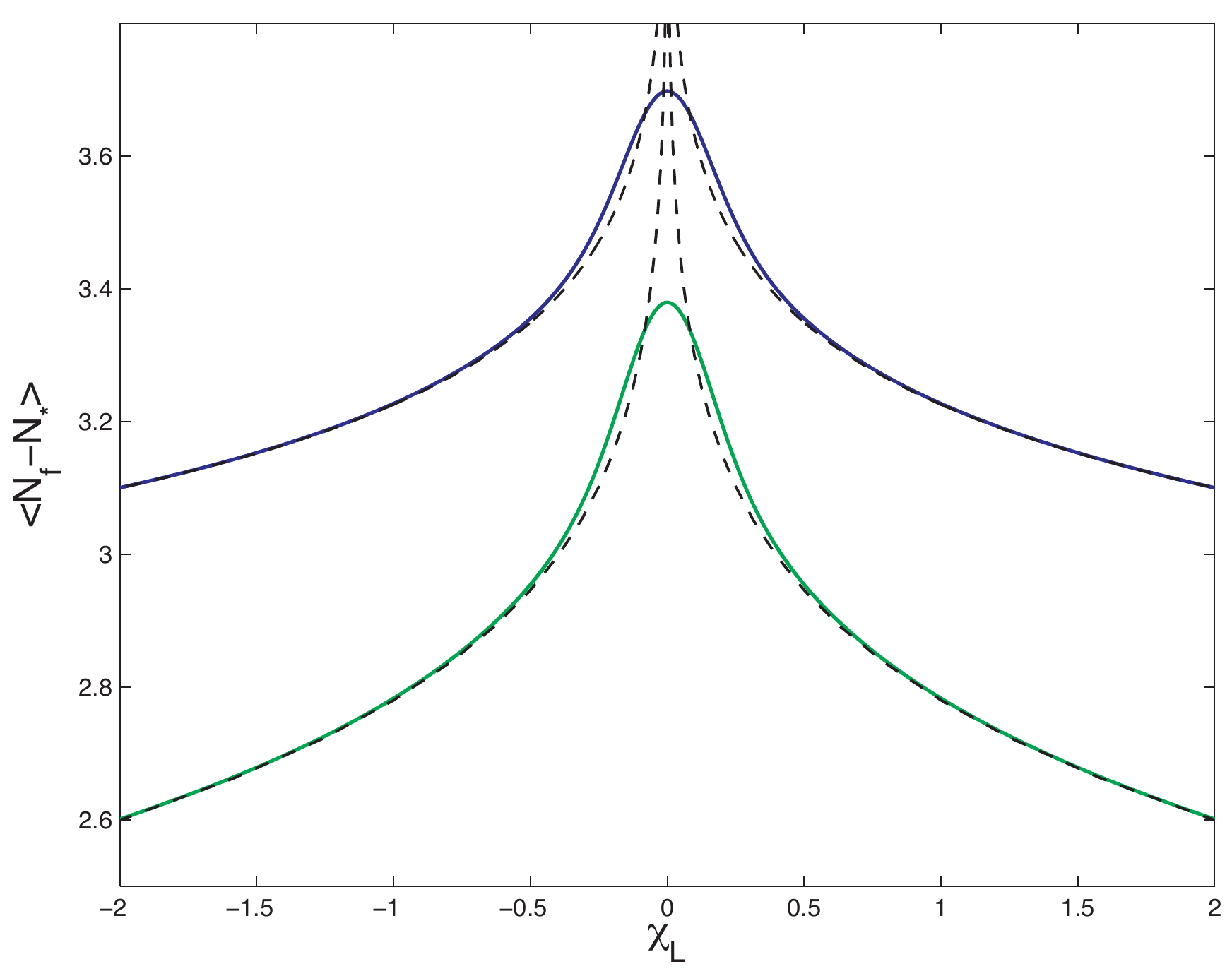}
\caption{ The integrated expansion from an initial time $\eta_*=-1$ to a final density corresponding to $\chi_f=10^5H_c$ as a function of the average value of the waterfall field, $\chi_L$, averaged on some super-Hubble scale, $L\gg |\eta_*|$. The upper dashed line shows the classical solution to the linearised equation (\ref{lineardeltachi}) in the long-wavelength limit, $k=0$, i.e., assuming a homogeneous field $\chi_*=\chi_L$. The lower dashed line shows the classical solution to the non-linear FRW equations (\ref{eqmphi}-\ref{Fried}). The solid lines shows the average expansion in each case integrated over a Gaussian distribution for the Hubble-scale field, $\chi_*$, with average value $\chi_L$ and variance $\sigma_*^2\simeq0.02H^2$. Both lines are obtained using $r=0.1$ and $\beta=100$.}
\label{nc-ns}
\end{figure}

The dashed line in Figure \ref{nc-ns} shows the classical expansion,
$N_f(\chi_*)=N(\chi_*\to\chi_f)$ from an initial value of the waterfall field
close to the transition, $\chi_*$, to a given final value,
$\chi_f$, for our chosen values of $r$ and $\beta$ using the linearised
equations of motion. In particular this shows the singular behaviour of
the classical solution for $N_f(\chi_*)$ about $\chi_*=0$.

%\begin{figure}
%\centering
%\includegraphics[width=0.9\textwidth]{/Users/zefonseca/Desktop/Hybrid/N-N*vslogchi=feta.pdf}
%\caption{ \scriptsize The vertical axis indicates the amount of expansion after the transition. The horizontal axes we plot the logarithm of the ration of the field value and its value at the transition. The dashed line indicates the analytic late time behaviour of $N-N_*$ as a function of $\chi/\chi_*$.}
%\label{n-ns}
%\end{figure}

% \begin{figure}
% \centering
% \includegraphics[width=0.9\textwidth]{Nc-Nvslog10chi_dchic.pdf}
% \caption{ \scriptsize The vertical axis indicates the amount of expansion after the transition. The horizontal axes we plot the logarithm of the ration of the field value at the instrability and end of inflation. The dashed line indicates the analytic late time behaviour of $Nc-N_*$ as a function of $\chi_*/\chi_c$.}
% \label{nc-nsvslog}
% \end{figure}

% \begin{figure}
% \centering
% \includegraphics[width=0.9\textwidth]{Nc-Nvschi.pdf}
% \caption{ \scriptsize The vertical axes shows the amount of expansion between the critical point and the beginning of the transition. In the horizontal axes we plot the value of the field at the transition.}
% \label{nc-ns}
% \end{figure}

Small scale quantum fluctuations play an essential part in the
transition as they are amplified by the tachyonic transition. The
variance of the $\chi$ field averaged on some scale,
% v2
$L=2\pi/k_L>(aH)^{-1}$ is given by integrating over all longer wavelengths
 \be
 \langle \delta\chi^2 \rangle_L = \frac{1}{a^2} \int^{k_L}_0 {\cal P}_u(k) {d\ln k}
 \,.
 \ee
On super-Hubble scales we have a steep blue spectrum
$\P_u(k)\propto k^3$ and hence $\langle \delta\chi^2 \rangle_L =
\P_u(k_L)/3a^2$. We see clearly from figure~\ref{pu2} that the
variance peaks on scales of order the Hubble scale at
the transition. Thus we need to include these Hubble-scale modes
in our estimate of the background expansion, $\langle N\rangle_L$, and the
perturbation, $\zeta=\delta \langle N\rangle_L$, due to longer wavelength modes.

% v2
We can always split a Gaussian random field, such as the initial $\chi$-field fluctuations at the transition, into a long-wavelength part and a short-wavelength part\footnote{
Note that we introduce an ultra-violet cut-off, $k_{UV}$, at small scales to avoid the UV-divergence of the fluctuations as $k\to\infty$. Modes with $k^2/a^2\gg 2\sqrt\lambda M^2$ do not experience any tachyonic growth and remain in the quantum vacuum state, hence we do not include them in $\chi_S$.}
:
 \be
\chi_*(x)
 = \chi_L(x) + \chi_S(x)
  = \int_0^{k_{\rm split}} d^3k \, \chi_k \, e^{-ikx} + \int^{k_{\rm UV}}_{k_{\rm split}} d^3k \, \chi_k \, e^{-ikx}\,.
 \ee
In this case we will identify $k_{\rm split}$ with the peak of the spectrum, around Hubble scale at the transition, $k_*$.
In a Gaussian field the long and short wavelengths modes are uncorrelated. Thus we may think of the Hubble-scale fluctuations as a statistically homogeneous distribution regularising the divergence of the classical solution for the long wavelength field $N(\chi_L)$, as illustrated in Figure~\ref{nc-ns}.

A precise calculation of the expansion through the transition
would require a full non-linear numerical simulation of the
inhomogeneous quantum fields on a lattice. We will instead make
two important simplifications which will enable us to make a
simple numerical estimate and which we expect to capture the essential
physics. Firstly we will treat the fluctuations as classical on
Hubble scales shortly after the transition. This becomes valid due
to the tachyonic growth of the mode function, but is only marginal
at the transition. And secondly we will use the separate universe
assumption, neglecting spatial gradients on these scales, which
again is only marginal at the transition. Modes leaving the Hubble
horizon at the transition will continue to grow relative to the
Hubble scale as inflation continues, but eventually will return
within the Hubble scale as inflation ends. The separate universe
assumption on these scales is only marginally valid and for a
limited period, but this is precisely the period we wish to study.

% v2
Using these assumptions we can then estimate the expansion, $N_f(\chi_*)$, in a Hubble-scale patch from an initial value, $\chi_*$, around the time of the transition. However in a much larger super-Hubble scale region with an average value for
the waterfall field, $\chi_L$, we sample many Hubble-scale patches and integrating over a large volume can be replaced by an integral over a Gaussian distribution for the initial values, $\chi_*$, given a local background value, $\chi_L$. This is an extension of the usual separate universe picture on large scales which incorporates small scale variance, $\sigma_*^2=\langle\chi_S^2\rangle_*$, uncorrelated with the long-wavelength field.
Thus we obtain
 \be
 \label{localNf}
 \langle N_f \rangle_{L} = \int_{-\infty}^\infty d\chi_* N_f(\chi_*) P(\chi_*|\sigma_*,\chi_L) \,,
 \ee
where $P(\chi_*|\sigma_*,\chi_L)$ is the Gaussian probability
distribution for the local value of the waterfall field in a
Hubble-size region
at time $t_*$ soon after $\phi=\phi_c$,
 \be
 \label{Gaussianchi}
 P(\chi_*|\sigma_*,\chi_L) = \frac{1}{\sqrt{2\pi}\sigma_*} \exp
 \left( - \frac{(\chi_*-\chi_L)^2}{2\sigma_*^2} \right) \,,
 \ee
given the average value on the larger scale, $\chi_L$, and the
variance of the field on Hubble scales, $\sigma_*^2\simeq P_u(k_*)/a_*^2$,
% v2
due to smaller scale fluctuations, $\chi_S(x)$.

The expression (\ref{localNf}) for $\langle N_f \rangle_{L}$ gives us the
expansion as a
% v2
smooth
function of the large-scale field
$\chi_L$, and thus the curvature perturbation can be
expanded about the background solution $\chi_L=0$ as a function of
the perturbation $\delta\chi_L$
 \be
 \zeta_\chi = \left. \frac{d\langle N_f \rangle_{L}}{d\chi_L} \right|_{\chi_L=0}
 \delta\chi_L + \frac12 \left.  \frac{d^2\langle N_f
 \rangle_{L}}{d\chi_L^2} \right|_{\chi_L=0} \delta\chi_L^2 +\ldots
 \,,
 \ee
where, given the Gaussian probability distribution
Eq.~(\ref{Gaussianchi}), we have
 \bea
 \label{dNdchi}
 \left.  \frac{d\langle N_f \rangle_{L}}{d\chi_L} \right|_{\chi_L=0} &=& \frac{1}{\sqrt{2\pi}\sigma_*}
 \int_{-\infty}^\infty
 d\chi_* \, N_f(\chi_*)\, \frac{\chi_*}{\sigma_*^2} \, \exp \left( - \frac{\chi_*^2}{2\sigma_*^2} \right)
 \,,\\
 \left. \frac{d^2\langle N_f \rangle_{L}}{d\chi_L^2} \right|_{\chi_L=0} &=& \frac{1}{\sqrt{2\pi}\sigma_*} \int_{-\infty}^\infty
 d\chi_* \, N_f(\chi_*)\, \frac{\chi_*^2-\sigma_*^2}{\sigma_*^4} \, \exp \left( - \frac{\chi_*^2}{2\sigma_*^2} \right)
 \,.
 \eea

{}From the symmetry of the system~(\ref{pot}) under $\chi\to-\chi$
we have $N_f(\chi_*)=N_f(-\chi_*)$ and hence we immediately see from
Eq.~(\ref{dNdchi}) that
 \be
 \left. \frac{d\langle N_f \rangle_{L}}{d\chi_L} \right|_{\chi_L=0} = 0 \,.
 \ee
Thus there is no linear contribution to the primordial curvature
perturbation from large-scale fluctuations in the waterfall field.
The leading order contribution to the primordial curvature
perturbation from fluctuations in the waterfall field on large
scales will be second order, and hence non-Gaussian
\cite{Enqvist:2004bk,Mulryne:2009ci}.

%In order to estimate $N_f(\chi_*)$ we will use numerical solutions
%for homogeneous fields with a linear approximation.

\subsection{Linearised solution}

Firstly we obtain numerical solutions for the classical evolution, $N_f(\chi_*)$, from an initial value of the waterfall field, $\chi_*$ when $\phi=\phi_c$, to a final value $\chi_f\gg \chi_*$ using the linearised equation of motion (\ref{lineardeltachi}) for $\delta\chi_k=\chi_*$, where we work in the long-wavelength limit and set $k=0$.
As expected, the classical evolution $N_f(\chi_L)$, shown by the dashed line in figure~\ref{nc-ns}, is singular in the limit $\chi_L\to0$. By contrast the solid line shows how the classical singularity at $\chi_L=0$ is regularised by the quantum dispersion of the local value, $\chi_*$, once we include smaller Hubble-scale modes, yielding a finite value for $\langle N_f\rangle_L$ as $\chi_L\to0$.

%In the linearised approximation the potential (\ref{pot}) is given by
%\be
% V (N,\chi) \simeq M^4 - \frac12 \beta H_c^2 \left( 1 - e^{-2rN} \right) \chi^2 \,,
% \ee
%and thus a uniform-$\chi$ hypersurface is approximately equal to a uniform-density hypersurface.

We can obtain a rough analytic estimate for $N_f(\chi_L)$ by noting that at sufficiently late times after the transition, $\phi\to0$ and we expect $\chi$ to have the late-time behaviour, $\chi\propto (-\eta)^{-s}$ given in Eq.~(\ref{lateutend}). Therefore, we expect $N_f$ and $\chi_f/\chi_*$ to be approximately given by
\be \label{n-n*}
N_f=\frac{1}{s}\ln \frac{\chi_f}{\chi_*}+\mbox{const.} \,.
\ee
Note that this assumes, $\phi\ll\phi_c$, by which point the linear approximation for $\chi$ is expected to have broken down. Nonetheless, substituting Eq.~(\ref{n-n*}) into Eq.~(\ref{localNf}) allows us to give an analytic estimate
 \be
 \label{<ncns>anal}
\langle N_f\rangle_L
 = N_* + \frac{1}{2s} \ln\bigg(\frac{2\sigma_*^2}{\chi^2_f} \bigg)+\frac{\ln2+\gamma_{\rm EM}/2}{\sqrt{2}s} -\frac{1}{2\sqrt{2}s}\frac{\delta\chi_L^2}{\sigma_*^2} + \mathcal{O}\left(\frac{\delta\chi^4_L}{\sigma_*^4} \right)
  \,,
 \label{nexpchil}
 \ee
where $\gamma_{\rm EM}=0.57721$ is the Euler-Mascheroni constant.

For our chosen parameters, $r=0.1$ and $\beta=100$, we have $\sigma_*^2\simeq \P_u/a_*^2 \simeq (H_c/2\pi)^2$, and we set $\chi_f=10^5H_c$. The analytic approximation (\ref{<ncns>anal}) then yields
 \be
 \langle N_f \rangle_L \simeq \mbox{const.}- 1.62 \frac{\delta\chi_L^2}{H_c^2} + \mathcal{O}\left(\frac{\delta\chi^4_L}{H_c^4} \right)
  \,.
 \ee
Performing the integration numerically in the linearised approximation one obtains
 \be
 \langle N_f \rangle_L \simeq \mbox{const.} - 4.62 \frac{\delta\chi_L^2}{H_c^2} + \mathcal{O}\left(\frac{\delta\chi^4_L}{H_c^4} \right)
  \,.
 \ee
So the analytic approximation~(\ref{n-n*}), gives a result roughly a factor of 3 away from the numerical result in the linear analysis.

\subsection{Non-linear FLRW solution}

As we have neglected gradient terms by setting $k=0$ in the linear equation for $\chi(N)$, we can in fact solve the full non-linear equations in the homogeneous limit, treating the local Hubble-scale patches as separate universes~\cite{Wands:2000dp} obeying the FLRW equations (\ref{eqmphi}--\ref{Fried}) in order to determine the coupled evolution of $\chi$ and $\phi$, and hence $N$, starting from $\chi=\chi_*$ when $\phi=\phi_c$. In this case we must also specify the energy scale of inflation. In Figure~\ref{nc-ns} we show the classical solution $N_f(\chi_L)$ as well as the averaged $\langle N_f\rangle_L$ including the small scale dispersion, where we have set $M\simeq10^{15}$~GeV and hence $H_c\simeq10^{11}$~GeV. The results are qualitatively similar to those obtained from the linear approximation. Non-linearities, in particular the $\chi$-dependent effective mass for the $\phi$ field, leads to a slightly more rapid transition, especially for larger initial values of $\chi_L$, and we find
 \be
 \langle N_f \rangle_L \simeq \mbox{const.} - 6.59 \frac{\delta\chi_L^2}{H_c^2} + \mathcal{O}\left(\frac{\delta\chi^4_L}{H_c^4} \right)
  \,.
 \ee

Note that in our non-linear, separate universe solutions we have calculated $N_f$ up to a final fixed density, and have verified that in practice this coincides with a final value for the waterfall field $\chi_f\simeq10^5 H_c$. For our chosen parameter values the universe is still inflating at this final time, with slow-roll parameter $\epsilon\equiv-\dot{H}/H^2\simeq0.001$. Thus our initial comoving Hubble-scale at $\eta_*$ is still larger than the final Hubble scale. But the slow-roll parameter is growing quickly and inflation ends soon afterwards ($N_{\rm end}\simeq N_f+0.4$).

Independently of the details of the calculation we see that the curvature perturbation due to large-scale perturbations in the waterfall field is second-order, and suppressed relative to the Hubble-scale variance of the field at the transition.
%
% v2
\begin{equation}
 \P_{\zeta_\chi}(k) \sim H_c^{-4} \P_{\delta\chi_L^2}(k) \,.
\end{equation}
where
\begin{equation}
 \P_{\delta\chi_L^2}(k) = \frac{k^3}{2\pi} \int d^3k' \frac{\P_{\delta\chi_L}(k')}{k^{\prime3}} \frac{\P_{\delta\chi_L}(|{\bf k}-{\bf k'}|)}{|{\bf k}-{\bf k'}|^{\prime3}} \,.
\end{equation}

Given the steep blue spectrum of the super-Hubble perturbations in the waterfall field, $\P_{\delta\chi_L}(k)=\P_u(k)/a^2\propto k^3$ (which implies a white spectrum in the
standard terminology since $P_{\delta\chi_L}(k)=2\pi^2\P_{\delta\chi_L}(k)/k^3\propto$ const.),
we thus conclude that the spectrum of the
resulting primordial curvature perturbation, $\zeta_\chi$, on super-Hubble scales
is also blue for $k\ll k_*$:
%  and is of order $(k_L/k_*)^3$.
\begin{equation}
 \P_{\zeta_\chi}(k) \sim H_c^{-4} \P_{\delta\chi_L}(k_*) \P_{\delta\chi_L}(k) \sim (k/k_*)^3\,.
\end{equation}
Assuming cosmological scales leave the Hubble-horizon around 40 e-folds before the end of inflation we have $(k_{\rm cmb}/k_*)^3\sim 10^{-54}$.

% \subsection{Non-linear, separate universe approximation}

\section{Conclusions}

We have estimated the primordial curvature perturbation produced by fluctuations in the waterfall field during hybrid inflation. We have calculated linear perturbations about the classical background trajectory, $\chi=0$, during slow-roll inflation and then studied how these affect the primordial curvature perturbation when the waterfall field is released and inflation comes to an end. To do this we have used a extension of the usual $\delta N$-formalism that identifies the primordial curvature perturbation with the perturbation in the local expansion on a uniform-density hypersurface, $\zeta=\delta N$.

The challenge for the standard $\delta N$ approach is that the homogeneous classical background solution, with smooth fields on scales much larger than the Hubble-horizon, fails to provide a good description of the dynamics. A classical solution which starts precisely in the false vacuum state stays there and inflation never ends. Thus even a tiny perturbation away from this background solution results in a huge apparent change in the local expansion, $N$. In fact the false vacuum is destabilised mainly due to fluctuations of the waterfall field on scales close to the Hubble-horizon at the end of inflation.

The coupling between the slow-rolling inflaton and the waterfall field that leads to the tachyonic instability also makes the waterfall field massive before the transition. Thus quantum fluctuations of the waterfall field are not amplified during inflation and remain in a quantum state even on super-Hubble scales until the critical value of the inflaton is reached. Once the tachyonic instability is triggered, there is an explosive growth of long-wavelength modes, but they retain the steep blue spectrum on super-Hubble scales. On the other hand much smaller scales, well inside the Hubble horizon at the transition are stabilised by spatial gradients and remain in their vacuum state. The resulting power spectrum for the waterfall field thus peaks on scales around the Hubble scale at the transition, as shown in Figure~\ref{pu2}, and these modes play an essential role in the dynamics that must be included when calculating the effect of large scale fluctuations in the waterfall field.

Thus we identify the primordial curvature perturbation due to perturbations in the waterfall field, $\delta\chi_L$
% v2 remove: averaged
on some large scale $L$, with the perturbation in the average expansion, $\langle N_f \rangle$, {\em including} fluctuations in the waterfall field on Hubble scales at the transition. We adopt a Gaussian distribution for the field whose average value in a region of size $L$ is $\delta\chi_L$, but whose variance is given by $\sigma_*^2=\P_u(k_*)/a^2$. We find two main results:
\begin{itemize}
\item
from the symmetry of the potential (\ref{pot}) under $\chi\to-\chi$ we see that the primordial curvature perturbation is independent of $\delta\chi_L$ at first order, and the curvature perturbation is second-order in $\delta\chi_L$.
\item
the spectrum of the primordial curvature perturbation due to fluctuations in the waterfall field on large scales is suppressed by a factor of order $(k_*L)^3$ which for cosmological scales is likely to be of order $10^{-54}$.
\end{itemize}
While we have considered the specific hybrid potential (\ref{pot}) and presented numerical solutions for specific parameter choices we believe the general conclusions will hold for all hybrid models in which the waterfall field is massive during slow-roll inflation and for which the end of inflation occurs due to a rapid tachyonic instability.

It is not surprising that the effect of small scale fluctuations must be taken into account when estimating the primordial curvature perturbation when the large-scale field is very close to zero. This is the case, for example, in the curvaton model where the curvature perturbation, $\zeta \sim \delta\sigma/\sigma$, could become very large when the background field, $\sigma$, is close to zero in some regions of the universe \cite{Linde:2005yw}. The apparent singular behaviour of the curvature perturbation in this case is regularised by smaller scale fluctuations in the curvaton field~\cite{Sasaki:2006kq}.

In our analysis we have considered only the early stages of the tachyonic instability. Eventually inflation ends and the coupled fields oscillate about the true vacuum $\phi=0$ and $\chi^2=2M^2/\sqrt\lambda$. Numerical simulations are required to study resonant particle production and other non-perturbative effects in this regime which goes beyond the scope of this paper. Resonance is often most efficient in long-wavelength modes, but fields which are massive 
% v2 during 
throughout 
slow-roll inflation necessarily give rise to a steep blue spectrum of perturbations and hence the large-scale power is suppressed with respect to smaller scales \cite{Liddle:1999hq,Barnaby:2006cq,Barnaby:2006km}.

\bigskip

{\em Note added:}
While this work was in progress, two papers appeared
\cite{Abolhasani:2010kr,Lyth:2010ch} which study the generation of large
scale curvature perturbations at the end of hybrid inflation from
complementary viewpoints and whose conclusions appear to be
consistent with our results.

\acknowledgements{The authors are grateful to Jim Cline, Hassan Firouzjahi, David Lyth, David Mulryne, David Seery and Daniel Wesley for discussions. JF is supported by Funda\c c\~ao para a Ci\^encia e a Tecnologia (Portugal), fellowships reference number SFRH/BD/40150/2007, and by the GCOE program at Kyoto University, ``The Next Generation of Physics, Spun from Universality and Emergence". DW is supported by the STFC. JF and DW are grateful to the Yukawa Institute for Theoretical Physics (YITP) for their hospitality where this work was initiated.
MS is supported in part by JSPS Grant-in-Aid for Scientific
Research (A) No.~21244033, and by JSPS
Grant-in-Aid for Creative Scientific Research No.~19GS0219.
We also acknowledge the workshop, `The non-Gaussian universe'
at YITP in March 2010, No.~YITP-T-09-05, during which part of this work
was done.}

\end{document}